\documentclass[runningheads,a4paper]{llncs}

\usepackage{times}
\usepackage{amssymb}
\usepackage{graphicx}
\usepackage{url}

\usepackage[T1]{fontenc}
\usepackage{amsmath}
\usepackage{bm}

\usepackage{graphicx}
\usepackage{booktabs}

\usepackage{fancyhdr}

\fancypagestyle{firstpage}{          
  \fancyhf{}                         
  \fancyfoot[r]{\footnotesize \it{This paper is accepted (camera‐ready) for TSD 2025.}}
}
\usepackage{color}

\newcommand{\keywords}[1]{\par\addvspace\baselineskip
\noindent\keywordname\enspace\ignorespaces#1}

\urldef{\mailsa}\path|{yi.zhong, cristian.tejedorgarcia, martha.larson}@ru.nl|
\urldef{\mailsb}\path|bas.bloem@radboudumc.nl|

\begin{document}

\title{RECA-PD: A Robust Explainable Cross-Attention Method for Speech-based Parkinson's Disease Classification}

\titlerunning{A Robust Explainable Cross-Attention Method for PD Classification}
%
\author{Terry Yi Zhong\inst{1} \and Cristian Tejedor-Garcia\inst{1} \and Martha Larson\inst{1} \and Bastiaan R. Bloem\inst{2}}


\authorrunning{Terry Yi Zhong et al.}

\institute{Centre for Language Studies, Radboud University, Nijmegen, the Netherlands 
\mailsa\\
\and
Center of Expertise for Parkinson and Movement Disorders, Radboud University Medical Center, Nijmegen, the Netherlands \\
\mailsb\\
}

\toctitle{} \tocauthor{}

\maketitle
\thispagestyle{firstpage}

\setcounter{footnote}{0}

\begin{abstract}
Parkinson's Disease (PD) affects over 10 million people globally, with speech impairments often preceding motor symptoms by years, making speech a valuable modality for early, non-invasive detection. While recent deep-learning models achieve high accuracy, they typically lack the explainability required for clinical use. To address this, we propose RECA-PD, a novel, robust, and explainable cross-attention architecture that combines interpretable speech features with self-supervised representations. RECA-PD matches state-of-the-art performance in Parkinson’s disease detection while providing explanations that are more consistent and more clinically meaningful. Additionally, we demonstrate that performance degradation in certain speech tasks (e.g., monologue) can be mitigated by segmenting long recordings. Our findings indicate that performance and explainability are not necessarily mutually exclusive. Future work will enhance the usability of explanations for non-experts and explore severity estimation to increase the real-world clinical relevance.

\end{abstract}

\section{Introduction}

 Parkinson's disease (PD) is the second most prevalent neurodegenerative disorder, affecting over 10 million individuals globally~\cite{1_ngo2022computerized,poewe2017parkinson}. In the early stages of the disease, patients may exhibit a range of speech-related abnormalities up to five years prior to the onset of marked motor symptoms~\cite{pdspeech1pinto2004treatments,pdspeech2rusz2013imprecise,mu2017parkinson}. It is estimated that between 70\% and 90\% of individuals with PD eventually develop vocal and speech impairments~\cite{hartelius1994speech,miller2007prevalence}.
Given that speech production involves the intricate coordination of cognitive and physiological mechanisms, speech-based approaches have emerged as promising automated, non-invasive, and cost-effective tools for PD detection~\cite{LisannevangelderenInnovativeSpeechBasedDeep2024}.
Nevertheless, despite these advantages, the unique challenges of the medical domain, coupled with increasingly stringent regulations surrounding Artificial Intelligence (AI), emphasize the critical need for explainability to ensure the safe and trustworthy deployment of such methods in real-world clinical practice~\cite{gambetti2025surveyhumancenteredevaluationexplainable}.

In recent years, a growing body of research has explored machine learning techniques for automatic, speech-based detection of PD \cite{conventionasquez-correaAutomaticEvaluationDysarthria2018,yuanyuanliuAutomaticAssessmentParkinsons2023}. This includes the use of deep neural networks that learn representations directly from raw speech signals~\cite{cnn1,cnnsonawaneSpeechbasedSolutionParkinsons2021}, and more recently, the incorporation of pre-trained self-supervised learning (SSL) models for speech representation~\cite{SSL4PR,mohamed2022self}.
Although these advancements have significantly improved classification performance, they have also reduced model interpretability, raising concerns about transparency and suitability for clinical deployment~\cite{InvestigatingEffectivenessExplainability2024a}.

To bridge this performance–explainability gap, recent research has explored explainable AI (XAI) approaches to speech-based PD detection, which can be broadly categorized into three main directions.
First, saliency-based XAI techniques have been applied to black-box models. For example,~\cite{InvestigatingEffectivenessExplainability2024a} showed that Mel spectrogram saliency maps offer faithful representations, but they still lack sufficient transparency for effective clinical interpretation.
Second, fully interpretable (white-box) models have been proposed~\cite{biomarkertelho2024speech}; however, these models currently fall short of achieving state-of-the-art (SOTA) performance.
Third, hybrid architectures integrate SSL speech representations with interpretable features using mechanisms such as cross-attention~\cite{lin2022cat}, as illustrated in~\cite{gimeno2025unveiling}.
This approach maintains competitive performance while offering explanations grounded in attention weights, and we adopt it as the base system for this paper. 
The approach presents two significant opportunities
for improvement by: (1) Refining the mathematical formulation of the cross-attention mechanism, thus taking a step forward in terms of consistency and technical robustness.
(2) Shifting the explanation to focus on the contribution of specific speech aspects, thus moving in the direction of increased potential for clinical usefulness.
These opportunities motivate our main research question (\textit{RQ}): \textit{To what extent can we develop a technically robust, explainable method for speech-based PD detection that provides more consistent and more clinically meaningful explanations?}





In this paper, we propose a novel method called \textbf{RECA-PD} (\underline{R}obust \underline{E}xplainable \underline{C}ross-\underline{A}ttention Method for Speech-based \underline{P}arkinson's \underline{D}isease
Classification), specifically designed to address the two shortcomings described above.
Our method begins by revisiting the cross-attention mechanism~\cite{hou2019cross} and correcting the softmax axis issue present in the implementation of the base method we build upon~\cite{gimeno2025unveiling}. 
To improve clinical explainability, we move a step forward by embedding the speech aspect categories into the model architecture through dedicated feature encoders.
Each encoder transforms the interpretable features corresponding to a specific speech aspect into distinct, learnable, interpretable tokens.
Finally, these aspect-specific tokens are subsequently used as \textit{Value} vectors in the cross-attention mechanism, replacing opaque SSL embeddings. As a result, the attention scores explicitly reflect the importance of interpretable speech aspects in the model's decision-making process, thereby enhancing both transparency and clinical relevance.

We evaluated our proposed method on the well-known in the literature PC-GITA dataset~\cite{PCGITAorozco2014new} following two established protocols~\cite{gimeno2025unveiling,SSL4PR},  and conducted ablation studies to assess the impact of each design component. 
Statistical analysis of these results demonstrates that our proposed method reduces the performance gap typically associated with using interpretable representations as the decision-making source, achieving performance comparable to a strong existing approach, i.e, ~\cite{gimeno2025unveiling}. Ablation studies further reveal that simply correcting the softmax axis issue in the base method yields significant improvements across multiple performance metrics.

To summarize, we outline the main contributions of this work as follows:
\begin{itemize}

\item We propose RECA-PD, a novel, robust, and explainable method for speech-based PD classification that delivers more clinically relevant explanations, with its implementation publicly available.\footnote{\scriptsize{\url{https://github.com/terryyizhongru/RECA-PD}}}

\item We perform comprehensive evaluations and ablation studies to assess RECA-PD and each design component.

\item Our results show that RECA-PD reduces the performance drop from using interpretable speech features, achieving results comparable to strong base methods.

 \end{itemize}

\section{Related Work}

Automatic speech-based PD detection has evolved significantly over the past decade. Early work focused on hand-crafted acoustic and prosodic features input into conventional classifiers~\cite{conventionasquez-correaAutomaticEvaluationDysarthria2018,yuanyuanliuAutomaticAssessmentParkinsons2023}, prior to the adoption of deep neural network architectures that learn representations directly from raw or lightly processed signals~\cite{cnn1,cnnsonawaneSpeechbasedSolutionParkinsons2021}. More recently, SSL-based speech embeddings~\cite{mohamed2022self} have set new performance benchmarks~\cite{SSL4PR,ParkCeleb_favaroUnveilingEarlySigns2024}. However, this emphasis on black-box models and accuracy gains has often come at the expense of explainability, limiting the clinical utility of these systems.


Due to the sensitive nature of the medical field, explainability is essential for the task of PD detection. In recent years, researchers have begun to pay significantly greater attention to the explainability of speech-based PD detection methods~\cite{InvestigatingEffectivenessExplainability2024a,biomarkertelho2024speech,gimeno2025unveiling}. ~\cite{InvestigatingEffectivenessExplainability2024a} systematically evaluates the effectiveness of saliency-based explainable AI approaches. The authors conclude that, although these methods provide faithful explanations via Mel spectrograms, further research is needed to make them more insightful for domain experts. This is because Mel saliency maps are difficult to interpret even for speech researchers, let alone for medical professionals in real-world scenarios. ~\cite{biomarkertelho2024speech} achieves meaningful explanation of PD detection by employing a fully white-box interpretable model, the Neural Additive Model~\cite{NAMagarwal2021neural}, built solely on interpretable features. However, there is a substantial drop in performance compared to other studies~\cite{SSL4PR,gimeno2025unveiling}, because such models can only capture first-order feature interactions and thus cannot model the co-relationships between input features~\cite{kimHigherorderNeuralAdditive2022}, which limits their fitting capacity. ~\cite{gimeno2025unveiling} takes this line of work a step further by fusing SSL representations with interpretable features via a cross-attention~\cite{lin2022cat} mechanism for PD classification. This method preserves classification performance, matching purely SSL-based methods, while producing visual explanations grounded in attention scores. Building on the foundational work of \cite{gimeno2025unveiling}, our method improves technical robustness and clinical explainability by refining the cross-attention formulation and anchoring explanations in specific speech aspects.

\section{Proposed Method}\label{sec:3}
This section is structured as follows. Section~\ref{sec:overall} provides a brief overview of the RECA-PD architecture. In Section~\ref{sec:token}, we describe the design of the feature encoder, which converts each aspect's raw speech feature set into a distinct, interpretable token. Section~\ref{sec:limitation-cross} revisits the cross-attention mechanism and discusses the opportunities for improvement of the base approach~\cite{gimeno2025unveiling}. Finally, in Section~\ref{sec:cross-attention}, we introduce our proposed cross-attention mechanism and emphasize its advantages in improving robustness and explainability.

\subsection{General Architecture}\label{sec:overall}
\begin{figure*}[ht!]
  \centering
  \includegraphics[width=0.9\linewidth]{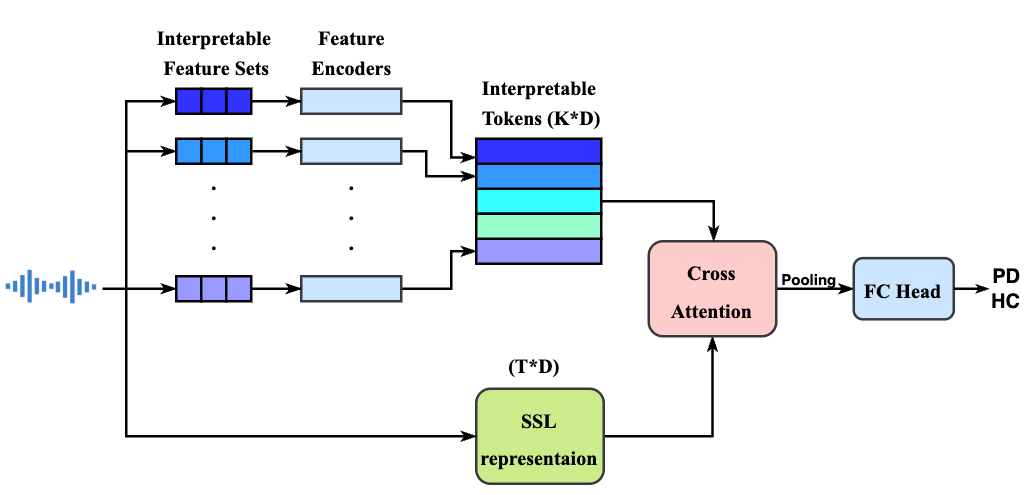}
  \caption{Overview of the proposed architecture of RECA-PD. $K$ denotes the number of speech aspect categories, $D$ is the dimension of the SSL representation, $T$ represents the number of time steps, FC stands for fully-connected network and PD and HC refer to Parkinson's disease and healthy control patients, respectively.}
  \label{fig:f1}

\end{figure*}

Figure~\ref{fig:f1} illustrates the overall architecture of RECA-PD. The interpretable features linked to each speech aspect are first encoded as distinct, interpretable tokens through separate encoders. Within the cross-attention module, these tokens serve as the sources of the \textit{Key} and \textit{Value} vectors, whereas pre-trained SSL embeddings provide the \textit{Query} matrix. The weighted sum produced by the attention mechanism is subsequently mean-pooled, and the resulting representation is forwarded to a fully connected layer for final classification.

\subsection{Interpretable Information Token}\label{sec:token}

We use the same 35 interpretable speech features with the base method~\cite{gimeno2025unveiling}, extracted using the DisVoice toolkit.\footnote{\scriptsize~\url{https://github.com/jcvasquezc/DisVoice}}
These features are categorized into four basic aspects of speech: articulation, glottal, phonation, and prosody, which represent the core elements of speech production affected by dysarthria and are familiar to speech experts.

These \(K\) feature sets (\(K=4\) in this study) are each encoded into a \(D\)-dimensional token by \(K\) separate feature encoders, each consisting of three linear layers interleaved with LayerNorm and Dropout. This design incorporates expert knowledge about the categories of different speech aspects into the architecture. Also, it prevents any interaction among the \(K\) aspects. This ensures that the weights of the subsequent attention mechanism, which relies on these tokens, exclusively and accurately reflect each aspect.

\subsection{Refining the Base Method's Cross-Attention Mechanism}\label{sec:limitation-cross}

Attention mechanisms~\cite{bahdanau2014neural} effectively model long-range dependencies and highlight salient patterns in sequential data, proving especially valuable in speech processing~\cite{chorowski2015attention}. Scaled dot-product attention is defined as:
{
\begin{equation}
\mathrm{Attention}(Q, K, V)
= \mathrm{softmax}\!\biggl(\frac{Q K^{\mathsf T}}{\sqrt{d_k}}\biggr)\,V \,
\end{equation}}
where $Q$, $K$, and $V$ are the \textit{Query, Key,} and \textit{Value} matrices, respectively, and $d_k$ is the key dimensionality used for scaling. These matrices are obtained via linear projections with learnable weights $W_Q$, $W_K$, and $W_V$. The softmax normalizes each row of the weight matrix into a probability distribution over the values, which are often referred to as attention scores. In cross-attention~\cite{hou2019cross}, the Query originates from a representation different from that of the key. Consequently, the attention scores describe relationships between two distinct representation spaces, and the scores serve as weights over $V$, which corresponds to one of those spaces.
\subsubsection{Shortcomings of the base method:} The base approach~\cite{gimeno2025unveiling} implements two different cross-attention mechanisms, \textit{Embedding} and \textit{Temporal}, both following:

\begin{equation}
\mathbf Q = \mathbf X^{\text{SSL}}_i \mathbf W_Q,\quad
\mathbf K = \mathbf X^{\text{inf}}_i \mathbf W_K
\text{ with }\mathbf W_K = \mathbf I,\quad
\mathbf V = \mathbf X^{\text{SSL}}_i \mathbf W_V.
\end{equation}

Embedding Cross-Attention repeats the informed features $T$ times so that $Q \in R^{D \times T}$ and produces shape scores $D \times F$; Temporal Cross-Attention instead repeats the informed features $D$ times, giving scores of shape $T \times F$.
%
Although the proposed idea is technically correct, we argue that the implementation and its resulting scores could be improved in terms of technical robustness and could be made more clinically meaningful.

For Embedding Cross-Attention, the resulting shape of the scores is $D \times F$. The softmax is done along the feature axis $F$ and then transposed to $F \times D$ so it can left-multiply $V$ ($D \times T$). However, transposing after softmax breaks the probability interpretation: after transposition, each row no longer sums to 1.
Furthermore, $Z = W^\top V$ becomes a non-convex sum of $D$ \textit{Value} vectors; its scale grows with $D$. Even when $D$ is fixed, this “sum” scale is disconnected from the rest of the model and can amplify or suppress activations, disrupting the numerical balance between network layers. 
For Temporal Cross-Attention, the same transpose happens after the softmax on $T \times F$, resulting in a real issue: because $T$ changes from batch to batch, the magnitude of every row of $W^\top V$ changes linearly with the audio duration, resulting in inconsistent scale across utterances and loss of comparability across different samples.

\subsection{Proposed Cross Attention mechanism}\label{sec:cross-attention}
In our proposed architecture, RECA-PD, the interpretable features associated with each PD-specific speech aspect are first encoded into separate interpretable tokens of size $D$. These $K$ aspect tokens serve simultaneously as the source of \textit{Key} and \textit{Value} ($K, V\in R^{K\times D}$) vectors, while the SSL embedding sequence provides the \textit{Query} ($Q\in R^{T\times D}$):
\begin{equation}
\mathbf Q = \mathbf X^{\text{SSL}}_i \mathbf W_Q,\quad
\mathbf K = \mathbf X^{\textbf{token}}_i \mathbf W_K,\quad
\mathbf V = \mathbf X^{\textbf{token}}_i \mathbf W_V\text{ with }
\mathbf W_V = \mathbf I.
\end{equation}

The resulting score matrix is therefore $W^\top V\in R^{T\times K}$; we apply the softmax along the $K$ axis, so no transposition is required, and each row remains a valid probability distribution over \emph{speech aspects}. Post-multiplying by $V$ ensures that the output matrix is a convex combination of the aspect embeddings, guaranteeing that it lies within the simplex spanned by these values. Since each row of $W$ sums to one, its coefficients can be interpreted directly as aspect-level importance weights driving the downstream classifier. This validates the claim that ``the attention weights directly correspond to speech aspects and drive the downstream classification.'' Moreover, by setting $W_V = \mathbf{I}$, we ensure that there is no interaction between different aspect tokens prior to the weighted sum. In other words, each aspect's contribution to $\mathbf{Z}$ is determined solely by its own attention weight, preserving the disentanglement of the $K$ speech aspects and enhancing the interpretability of the explanation scores.

Even after correcting the score matrix computation by transposing it before applying the softmax operation along the $D$ or $T$ dimension to preserve probabilistic semantics, the explanations of the baseline approach are not sufficiently clinically relevant. Specifically, its embedding-level cross-attention addresses the question: \textit{``To what extent does each interpretable speech feature attend to the various latent SSL channels?''} While this insight benefits speech-technology researchers, it provides minimal utility for clinicians.
Similarly, the temporal cross-attention mechanism aims to identify \textit{``when a feature is salient''}, but the softmax applied over time constrains attention weights to sum to one across all time steps, preventing an assessment of a feature’s absolute importance in the overall decision.
In contrast, our proposed design delivers more clinically actionable insights by directly identifying \textit{which speech aspect most significantly influences the model’s final prediction.}

\section{Experimental Procedure}


\subsection{Dataset}We use the widely recognized PC-GITA dataset~\cite{PCGITAorozco2014new} for our experiments, following the base approach~\cite{gimeno2025unveiling}.
The dataset comprises 100 participants: 50 individuals with PD and 50 healthy controls (HC) with ages ranging from 31 to 86 years. Each group consists of 25 males and 25 females. All PD participants were clinically diagnosed by a neurologist.  
The dataset contains six types of speech tasks with the following numbers of utterances: 15 sustained vowels, 6 diadochokinetic (DDK) exercises, 25 isolated words, 10 read sentences, 1 read passage, and 1 spontaneous monologue.
All recordings were conducted while patients were in the "on" state of their medication. Speech data was collected in Colombian Spanish within a soundproof booth. 

\subsection{Evaluation Methods}\label{sec:eval}
We evaluate the classification performance of the proposed method using two methods. First, following the base approach protocol described in~\cite{gimeno2025unveiling}, we train and test the model on each task independently, reporting only the F1-score for comparison (Table~\ref{tab:task_f1_scores_transposed}). Because the monologue recordings are both significantly longer and contain fewer utterances, we segmented each into ten shorter clips, creating a Split-Mono dataset for more comprehensive analysis.
Second, to provide a more comprehensive assessment using clinically relevant metrics, we adopt the protocol from~\cite{SSL4PR}, which involves training and testing on a combined set of DDK exercises, read sentences, and spontaneous monologues. 
In this setting, we report accuracy, precision, F1-score, AUC-ROC, sensitivity, and specificity (Table~\ref{tab:performance_comparison}). For both evaluation strategies, we use the same speaker-independent nested cross-validation splits as in the referenced methods.
\subsection{Experimental Settings}
We replicate the results of the base method (M1) using the official implementation.\footnote{\scriptsize\url{https://github.com/david-gimeno/interpreting-ssl-parkinson-speech}} To facilitate a fair comparison with the base approach, we conduct two ablation experiments, resulting in methods M2 and M3. M2 addresses the softmax axis issue in M1 during attention computation (FixSoftmax), as detailed in Section~\ref{sec:limitation-cross}. Method M3 builds upon M2 by replacing the source of the \textit{Value} in the cross-attention mechanism—originally SSL embeddings—with interpretable speech features (InterpretableValue). Finally, M4—our proposed RECA-PD—builds upon M2, but replaces the raw input features used as \textit{Value} vectors with interpretable tokens extracted from the feature encoder (InterpretableToken). Table~\ref{tab:models} summarizes all the methods used in our experiments, and Section \ref{sub:implementation} provides specific implementation details for each one of them.

\begin{table}[ht!]
\centering
\small
\vspace{-0.25cm}

\caption{Description of the methods used in our experiments}
\begin{tabular}{llp{10cm}} 
\toprule
\textbf{Method} & & \textbf{Description} \\
\midrule
\textbf{M1} & & Base method~\cite{gimeno2025unveiling}. \\
\textbf{M2} & & M1 + FixSoftmax: fixes the softmax axis issue. \\
\textbf{M3} & & M2 + InterpretableValue: uses interpretable features as the \textit{Value} vectors source in cross-attention. \\
\textbf{M4} & & Proposed RECA-PD (M2 + InterpretableToken): uses interpretable tokens from the feature encoder as \textit{Value} vector instead of raw features. \\
\bottomrule
\end{tabular}
\label{tab:models}
\vspace{-0.25cm}
\end{table}

All experiments are conducted on a single A10 GPU, with the training configuration matching the base approach.\footnotemark[3] Each experiment is run five times with different random seeds to ensure reliability following~\cite{gimeno2025unveiling,zhong2025evaluatingusefulnessnondiagnosticspeech}. We first compute the mean and standard deviation of all metrics across all folds and then report the average of these metrics across five runs.

\subsection{Implementation Details}\label{sub:implementation}

We implement M2, M3, and our proposed RECA-PD (M4) based on the official implementation of the base method.\footnotemark[3] In M2, we address the softmax axis issue by first transposing the score matrix and then applying the softmax function along the SSL-channel dimension \(D\) for the \textit{embedding} head and along the temporal dimension \(T\) for the \textit{temporal} head, rather than along the feature dimension \(F\), as described in Section~\ref{sec:limitation-cross}. M3 builds upon M2 by replacing the \textit{Value} matrix of the two attention mechanisms—originally the SSL embeddings—with the corresponding repeated raw informed speech features explained in Section~\ref{sec:limitation-cross}.
For the proposed RECA-PD (M4), we only keep one attention mechanism, which replaces the sources of the \textit{Key} and \textit{Value} vectors with the interpretable tokens produced by the feature encoder and remove the linear projection preceding the \textit{Value} vectors, as described in Section \ref{sec:cross-attention}. The feature encoder consists of three linear layers: the first one has input dimensionality equal to the number of input features and output dimensionality of 128, followed by layers of size 512 and 1024, each interleaved with LayerNorm and a dropout rate of 0.1. All other model modules remain unchanged from the base method.

\section{Results and Discussion}

\subsection{Classification Performance}
This section presents and interprets the performance results of the proposed methods, with a focus on both task-specific evaluation and overall performance on multiple metrics. The goal is to assess whether improvements in explainability can be achieved without compromising accuracy.
%

Table~\ref{tab:task_f1_scores_transposed} summarizes method performance per task using the base approach evaluation protocol, while Table~\ref{tab:performance_comparison} presents a comprehensive comparison, including all evaluation metrics and Wilcoxon signed-rank test~\cite{woolson2005wilcoxon} significance when comparing each metric to the corresponding metric of the
preceding method in the list (for M1, comparison is to M4).
In these two tables, it is shown that M4 performs on par with M1 and nearly matches M2, while being specifically designed to enhance explainability and significantly outperforming M3. This demonstrates that our proposed approach 
successfully balances accuracy and explainability, 
marking an important step forward in explainable PD detection.

\begin{table*}[ht!]
\centering
\small
\caption{F1 scores (\%) across tasks for method M1 to M4.}
\begin{tabular}{lcccc}
\hline
\small
\textbf{Task}      & \textbf{M1~\cite{gimeno2025unveiling}} & \textbf{M2} & \textbf{M3} & \textbf{M4 (Proposed)} \\ 
\hline
Vowels             & 65.7$\pm$7.0      & 62.0$\pm$6.0        & 62.2$\pm$5.0      & 62.8$\pm$8.2      \\ 
Words              & 73.1$\pm$4.6      & 74.4$\pm$4.1        & 65.5$\pm$5.2      & 70.3$\pm$5.6      \\ 
DDK                & 77.0$\pm$5.6      & 77.0$\pm$6.0        & 67.7$\pm$4.8      & 74.3$\pm$5.7      \\ 
Sentences          & 77.7$\pm$8.2      & 79.6$\pm$4.6 & 70.3$\pm$4.0      & 78.0$\pm$6.1      \\ 
Read               & 76.1$\pm$12.7     & 81.3$\pm$5.9 & 59.5$\pm$9.6      & 75.4$\pm$7.0      \\ 
Monologue          & 76.9$\pm$10.5     & 77.2$\pm$7.4        & 58.4$\pm$10.1     & 66.5$\pm$5.6      \\ 
Split-Mono         & 77.6$\pm$5.5      & 76.4$\pm$5.6        & 66.3$\pm$4.5      & 73.8$\pm$7.8      \\ 
\hline
\end{tabular}
\label{tab:task_f1_scores_transposed}
\end{table*}


\begin{table*}[ht!]
\centering
\small
\caption{Performance comparison of different methods on different metrics. * indicates $p<0.05$ (Wilcoxon signed‐rank test) when comparing each metric to the corresponding metric of the preceding method in the list (for M1, comparison is to M4).}
\begin{tabular}{lcccccc}
\hline
\small
\textbf{Method} & \textbf{Acc.} & \textbf{F1-Score} & \textbf{Prec.} & \textbf{AUC} & \textbf{Sens.} & \textbf{Spec.} \\
\hline
\small
\textbf{M1~\cite{gimeno2025unveiling}}   
  & 77.9±7.5 & 77.5±7.9 & 79.2±8.7 & 83.0±9.6 & 76.5±9.8 & 79.2±10.2 \\
\textbf{M2}
  & 79.4±7.8 & 79.2±8.1* & 80.1±8.4 & 84.5±9.1* & 78.8±9.4* & 80.1±9.1 \\
\textbf{M3}
  & 76.0±8.4* & 75.9±8.5* & 76.8±9.5* & 82.3±9.7* & 75.6±9.9* & 76.4±11.2* \\
\textbf{M4 (Proposed)}  
  & 77.9±7.6* & 77.3±8.0* & 79.8±9.2* & 83.2±10.3 & 75.6±9.7 & 80.2±9.7* \\
\hline
\end{tabular}
\label{tab:performance_comparison}
\end{table*}

Regarding the ablation studies, M2 achieves significantly better performance over most metrics, surpassing the base approach by addressing the softmax axis issue. While M2 demonstrates the upper bound of performance in this study, M4 provides a more explainable alternative with competitive accuracy. In contrast, Method M3, which relies exclusively on interpretable features as \textit{Value} vectors in the cross-attention mechanism to enhance explainability, suffers a significant performance drop across all tasks and metrics. This aligns with previous findings and suggests that interpretable features alone lack sufficient representational power for competitive PD detection. These results further emphasize the value of the design in M4, which results in a better balance of explainability and representational richness.

Both M3 and M4 perform worse on the Monologue task compared to M2. Unlike M2, these methods do not use SSL embeddings as \textit{Value} source in the cross-attention mechanism, a factor that has been shown to be beneficial for spontaneous speech tasks~\cite{gimeno2025unveiling}. Besides this factor, the Monologue dataset differs by having approximately ten times fewer training samples with much longer recordings, also likely contributing to the performance gap. To further investigate this issue, we conducted an additional experiment in which each monologue recording was segmented into 10 shorter audio clips (Split-Mono). This adjustment ensures the dataset size and audio length align with those of other tasks. As shown in the last row of Table~\ref{tab:task_f1_scores_transposed}, the previously observed performance drop on the Monologue task was greatly alleviated when training on these segmented recordings. This finding suggests that the original performance degradation was primarily due to limited sample size and longer audio duration, rather than the absence of SSL representations alone. This result further supports the robustness of our approach and contributes to answering our \textit{RQ} by clarifying the sources of variability in model performance.

%
Overall, M4 demonstrates a strong balance between explainability and performance, making it a promising method for explainable PD detection,  particularly when it is necessary to balance explanation quality with predictive accuracy.


\subsection{Discussion of the Generated Explanations}
\vspace{-0.1cm}

\begin{figure*}[ht!]
  \centering
  \includegraphics[width=0.85\linewidth]{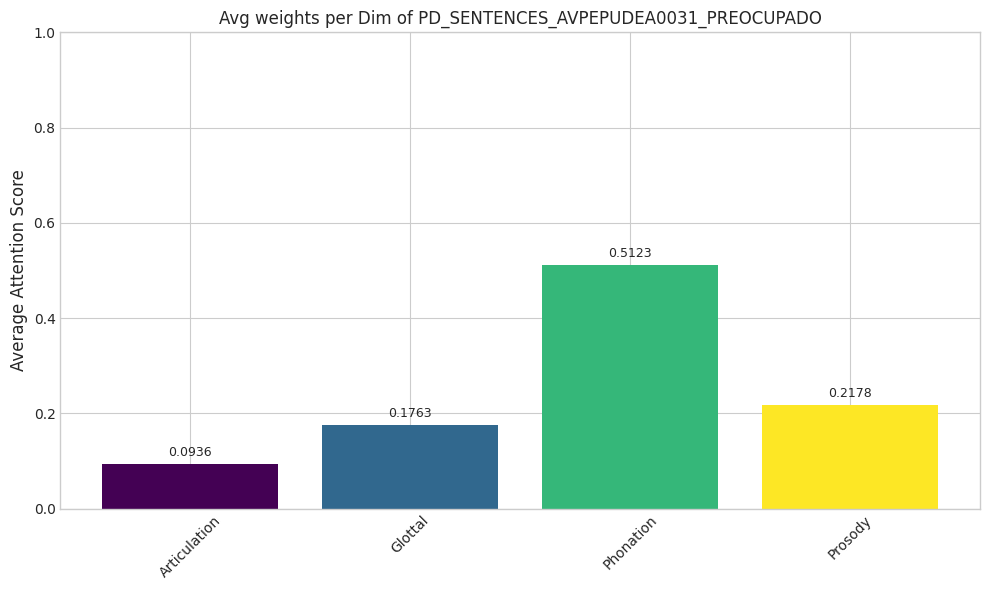}
  \vspace{-0.2cm}
  \caption{A visual example of the explanation generated by the proposed method.}
  \label{fig:f2}
\end{figure*}

Figure~\ref{fig:f2} shows a visual explanation for a randomly selected utterance generated by our proposed method under the second evaluation protocol (Section~\ref{sec:eval}). We use the same plotting conventions as the base method~\cite{gimeno2025unveiling} to illustrate the explanation scores across the four fundamental speech dimensions introduced in Section~\ref{sec:token}.

The scores produced by our approach offer two significant advantages. 
First, they demonstrate improved technical robustness. As discussed in Section~\ref{sec:limitation-cross} and evident from the examples provided in the original paper describing the base method, the explanation scores across different dimensions in the base approach vary considerably (e.g., ranging from 0 to 25 shown in Figure 5 in~\cite{gimeno2025unveiling}), and the sums of these scores are inconsistent across samples, making the weights difficult to compare reliably. By correcting the softmax axis in our approach, the resulting scores form a convex combination, where each row sums to one, providing a consistent probability distribution over the speech aspect categories. 
Second, the scores derived from our method are more clinically meaningful. By introducing interpretable speech aspect tokens as the \textit{Value} vectors in the cross-attention module, the model is constrained to route information through these tokens, thereby anchoring the learned similarities to predefined clinically relevant aspects. In contrast, the base method's cross-attention mechanism tends to highlight SSL channels that already contain task-relevant signals, which results in explanations that focus more on the internal SSL embeddings rather than on the clinically significant factors influencing the model's decisions.

\vspace{-0.25cm}
\vspace{-0.25cm}

However, our current explanation approach still exhibits certain limitations. The generated explanations may remain too abstract for some potential users. While the four dimensions reflect the fundamental components of speech production affected by dysarthria and are well understood by speech-language experts, the intended users of the PD classification system, such as general practitioners, neurologists, or patients conducting self-assessments, may not be familiar with these specialized concepts. This could limit the system's usability in broader clinical contexts. Additionally, although the explanation scores indicate the relative importance of each speech aspect in the model's decision-making process, they do not necessarily convey information regarding the severity of these aspects, which would be beneficial for practical use.

\section{Conclusion}
In this work, we developed a novel, robust, and explainable method for speech-based PD detection, RECA-PD, by integrating interpretable speech features with SSL representation in a cross-attention architecture.
We answered our \textit{RQ} affirmatively: RECA-PD can indeed be designed to provide reliable and meaningful explanations without compromising classification performance. Notably, while method M2—using SSL representations as the \textit{Value} matrix source—demonstrates advantages in overall performance, our proposed RECA-PD (M4) employs interpretable tokens as the \textit{Value} vector, offering a better trade-off between explainability and performance.
Our findings highlight the significance of hybrid architectures and proper normalization of attention scores. 
However, certain limitations persist; for instance, the speech-aspect categories of the explanation may be unfamiliar to clinicians, potentially reducing clinical utility.
Future work will include redefining the speech‐aspect categories using PD‐specific domain knowledge to make the explanations more accessible to a broader audience, analyzing attention scores across diverse speech tasks, and extending to other neurodegenerative diseases.

\vspace{-0.1cm}
\begin{credits}
\subsubsection{\ackname} 
This work is part of the project Responsible AI for Voice Diagnostics (RAIVD) with file number NGF.1607.22.013 of the research program NGF AiNed Fellowship Grants, which is financed by the Dutch Research Council (NWO). This work used the Dutch national e-infrastructure with the support of the SURF Cooperative using grant no. EINF-10519. 

\vspace{-0.2cm}
\subsubsection{\discintname}
The authors have no competing interests to declare that are
relevant to the content of this article. 
\end{credits}
\vspace{-0.1cm}
%
%
%
\bibliographystyle{splncs04}
\bibliography{mybibnew}
%





\end{document}